\documentstyle[mbf,epsf]{ptptex}

\title{Phase transition of color-superconductivity and cooling behavior of quark stars
}
 
\author{Keiji {\sc Yamaguchi}, Masaharu {\sc Iwasaki}$^{*,}$ and Osamu {\sc Miyamura}
}
 
\inst{Department of Physics, Hiroshima University, Higashi-Hiroshima 739-3526
\\
$^*$ Department of Physics, Kochi University, Kochi 780-8520
}
 
\recdate{
}
 
\abst{
We discuss the color-superconductivity and its effect on the cooling behavior of strange quark stars. The neutrino emissivity and specific heat of quark matter are calculated within the BCS theory. In the superconducting phase, the emissivity decreases and causes suppression of the cooling rate. It is shown that the phase transition leads to a sudden discontinuous suppression of the cooling rate in cooperation with the specific heat.
}
 
\begin{document}
 
\maketitle
 
\section{Introduction}
Recently many authors have investigated the color-superconductivity in quark matter.\cite{rf:1} \ Although there is no evidence of its existence, it is pointed out that there might be exist in the core of neutron stars and in the central region of high-energy heavy-ion collisions.
 
What physical quantities must we observe in order to show the existence of color-superconductors? In this paper, we discuss physical phenomena, which are characteristic of the color-superconductivity. To this end, we will focus our attention on the cooling behavior of the hot strange quark stars.\cite{rf:2} \ The recent works show that the quark matter is transformed from Fermi gas into superconducting fluid with the decrease of the temperature; there occurs a phase transition of second order at a critical temperature.\cite{rf:3} \ Therefore it is expected that some physical quantities have singularity when the hot quark matter is cooled down. This singularity may be a possible signal for the existence of color-superconductors.
 
The cooling phenomenon of quark stars has been studied for many years by several authors.\cite{rf:4} \ According to their considerations, only the week interaction can release the energy from such a high-density quark matter; the simplest possible process is neutrino emission from the quark matter and in particular the pair of ${\it d}$ quark $\beta$ decay reaction is dominant (URCA):
\begin{eqnarray}
\left\{ \begin{array}{rl}
d\to u+e^{-}+\bar{\nu}_e , \\
u+e^{-}\to d+\nu_e .
\end{array} \right.
\end{eqnarray}
The neutrino energy loss at an interval $dt$ for via this process is denoted by $C_{m}dt$. The constant $C_{m}$ is called neutrino emissivity. On the other hand, noting that it is expressed by $C_{V}dT$ with the specific heat $C_V$ of the quark matter, we obtain a differential equation determining the temperature as a function of time,
\begin{equation}
\frac{dT}{dt}=-\frac{C_m(T)}{C_V(T)}.
\end{equation}
This equation describes the cooling behavior of the quark matter.

Recently much efforts have been made in order to study the influence of the color-superconductivity on the cooling behavior.\cite{rf:5} \ They are, however, restricted to the low temperature regions and dealing with mainly the specific heat. It is the purpose of this paper to investigate the effect of the neutrino emissivity on the cooling behavior around the critical temperature.

In the next section, the mean field (BCS) theory for the color-superconductivity is developed in such a way that it is suitable for our present case of finite temperature mean field theory. Then we calculate the neutrino emissivity and specific heat of quark matter in \S 3. Finally (\S 4) numerical calculations are carried out for a typical quark matter and some comments are given.

\section{BCS theory at finite temperature}
First let us develop the BCS theory in the three-flavor quark system at finite temperature. The Lagrangian is assumed to be given by

\begin{equation}
{\cal L}=\overline{\Psi}(i\gamma \cdot \partial+\mu \gamma^{0})\Psi
            +g\sum_{a}(\overline{\Psi} \gamma^{\mu} \lambda^{a}\Psi)(\overline{\Psi} \gamma_{\mu} \lambda^{a}\Psi),
\end{equation}
where $\mu$ is a chemical potential and $\lambda^{a}$ denotes the color SU(3) matrix. This effective Lagrangian comes from the one-gluon exchange interaction with infinite gluon mass due to the many-body effect in the medium. To study pair correlation, it is convenient to use the Fierz transformation \cite{rf:5},

\begin{equation}
{\cal L}=\overline{\Psi}(i\gamma \cdot \partial+\mu \gamma^{0})\Psi
            +\frac{2}{3}g{\sum_{a,b}}'(\overline{\Psi} \gamma^{5}C \lambda^{a}\Lambda^{b}\overline{\Psi}^{t})
   (\Psi^{t}C^{-1} \gamma^{5} \lambda^{a}\Lambda^{b}\Psi),
\end{equation}
where $\Lambda^{b}$ denotes the flavor SU(3) matrix. Here we have left only the most attractive terms: spin-singlet, color-antisymmetric and flavor-antisymmetric ($a,b=2,5,7$) terms.

The partition function of our system is given by the following functional integral,

\begin{equation}
{\rm Z}={\rm Tr \, exp}(-\beta \hat{H})=\displaystyle \int{\cal D}\overline{\Psi} {\cal D}\Psi \,{\rm exp}(-\displaystyle \int d^4x {\cal L}),
\end{equation}
with the use of $\beta\equiv 1/T$ and $d^4x = d\tau d^3x$. Here let us introduce auxiliary field $\varphi(x)$ in terms of the following identity: 

\begin{equation}
1=\displaystyle \int{\cal D}\varphi^{*} {\cal D}\varphi \,{\rm exp}(-\kappa^2 \displaystyle \int d^4x |\varphi_{\rho}(x)|^2).
\end{equation}
Using this equation, we have chosen the constant $\kappa$ in the above function in such a way that the four-Fermi interaction term is eliminated,

\begin{eqnarray}
Z&=&\displaystyle \int{\cal D}\overline{\Psi} {\cal D}\Psi {\cal D}\varphi^{*} {\cal D}\varphi \,{\rm exp}
[-\displaystyle \int d^4x\{\overline{\Psi}(i\gamma \cdot \partial+\mu \gamma^{0})\Psi \nonumber \\
& &+(\overline{\Psi} \gamma^{5}C \lambda^{a}\Lambda^{b}\overline{\Psi}^{t}\varphi_{\rho}
      +\Psi^{t}C^{-1} \gamma^{5} \lambda^{a}\Lambda^{b}\Psi\varphi^*_{\rho})
   +\kappa^2|\varphi_{\rho}|^2 \}], 
\end{eqnarray}
where the color-flavor matrix $T^{\rho}$ is defined by $\lambda^a \otimes \Lambda^b$ . From the above equation it is clear that the auxiliary field is equivalent to a wave function of Cooper pairs.

Here we take a mean field approximation (BCS theory), assuming that the field $\varphi(x)$ is a constant,

\begin{equation}
\varphi_{\rho}(x) = \delta_{a,2}\delta_{b,2} \varphi_{0}.
\end{equation}
This expression means that our Cooper pairs are {\it u-d} spin-singlet with the momenta $\pm {\mbf k}$. The reason that we have taken {\it u-d} pairing is that the mass of the {\it s} quark is heavier than those of {\it u,d} quarks.

To carry out the integrations, we expand the Fermi field $\Psi$ into Fourier series:

\begin{equation}
\Psi(x)=\displaystyle \sum_n \displaystyle \sum_{{\mbf p}}\displaystyle \sum_{s} b_{p,s}u({\mbf p},s){\rm exp}\{i({\mbf p}\cdot {\mbf x}-\omega_n \tau)\},
\end{equation}
where the Matsubara frequency $\omega_n$ represents $\equiv (2n+1)\beta^{-1}$ and $s$ ($\pm 1$) is an eigenvalue of the quark helicity. Substituting this equation into the partition function, it can be represented as the following matrix form:
\begin{equation}
-S = \beta \displaystyle \sum_{p}\left[
\left(
\begin{array}{cc}
b^{\dag}_{p,1} & b^{t}_{-p,1} 
\end{array}
\right)
\left(
\begin{array}{cc}
 i\omega_n -\zeta_{p} & \Delta_{0} \\
 \Delta^{*}_{0} & i\omega_n +\zeta_{p}
\end{array}
\right)
\left(
\begin{array}{c}
b_{p,1}\\
b^{*}_{p,-1} 
\end{array}
\right)
+\kappa^2 |\varphi_{0}|^{2} \right],
\end{equation}
where the usual gap parameter has been defined by $\Delta_{0} \equiv -2 \varphi_{0} T^2 (T^2 \equiv \lambda^{2} \otimes \Lambda^{2}) $. This Cooper pair is antisymmetric in spin, color and flavor spaces independently.

Carrying out the functional integrals of the quark fields, we obtain

\begin{equation}
Z={\rm N}\, {\rm exp}\left[\displaystyle \sum_{p} {\rm Tr}(\log \beta^2(-\omega^{2}_{n}-\hat{E_{p}}^2))-\beta \kappa^2 
\frac{\Delta_{0}^2}{4}\right],
\end{equation}
where $\hat{E_{p}}$ is the energy matrix of the quasi-particle and given by
\begin{equation}
\hat{E_{p}}^2 = \zeta_{p}^2 + \Delta_{0}^2 (\lambda^2 \otimes \Lambda^2)^2.
\end{equation}
The characteristic of the quasi-particle is the following; all the {\it s} quarks are gapless and have not pair correlation, on the other hand the {\it u,d} quarks with the color indices 1 and 2 have a finite gap and the remaining particle is gapless.\cite{rf:6}

From the partition function (7), one can easily calculate the thermodynamical potential,

\begin{eqnarray}
\Omega &=& -\beta^{-1} \log Z = -4 \displaystyle \sum_{{\mbf p}}(E_{p}+2\beta^{-1} \log(1+{\rm exp}(-\beta E_{p})) \nonumber \\
& &-5 \displaystyle \sum_{{\mbf p}}(\xi_{p}+2\beta^{-1} \log(1+{\rm exp}(-\beta \xi_{p}))
+\frac{1}{4}\kappa^2 \Delta^2_{0}.
\end{eqnarray}
The gap parameter can be determined through the stationary condition of the thermodynamical potential,

\begin{equation}
\frac{\partial \Omega}{\partial \Delta_{0}} = -4 \displaystyle \sum_{{\mbf p}} \frac{\Delta_{0}}{E_{p}}
(1-2n_{p})+\frac{\kappa^2}{2}\Delta_0 = 0,
\end{equation}
with $n_{p} \equiv (1+ {\rm e}^{\beta E_{p}})^{-1}$, the distribution function of the quasi-particle.

\section{The neutrino emissivity and the specific heat}
Now we are in a position to discuss the cooling behavior by using the above results \cite{rf:4,rf:8}. First let us calculate the specific heat of the quark matter. To this end, we must derive entropy of the quark matter, which is written as

\begin{equation}
S=-\left(\frac{\partial \Omega}{\partial
T}\right)_{V,\mu}=-8\sum_{p}\left[n_{p}\log n_{p} +(1-n_{p})\log (1-n_{p})
\right].
\end{equation}
Here we have omitted the contribution of anti-quarks for the sake of brevity. They will be taken into account in the numerical calculations in the next section. From the entropy, we get the specific heat as usual way:

\begin{equation}
C_{V}=T\left( \frac{\partial S}{\partial T} \right)_{V,\mu}.
\end{equation}
Note that another specific heat $C_p$ is nearly equal to the $C_{V}$ in the Fermi gas as is well known in statistical physics. The entropy does not contain the derivative of the gap variable by virtue of the gap equation (14). Hence it is continuous function of $T$ across the critical temperature. On the other hand, the specific heat contains the derivative of the gap variable so that it is discontinuous as a function of the temperature, which is well known in electron superconductors. The sudden enhancement of the specific heat comes from the formation of Cooper pairs. Because a piece of heat applied to the superconductor is expended on the destruction of Cooper pairs instead of thermal motion.
 
Next let us consider the neutrino emissivity. It was calculated in the quark matter in the normal phase long years ago.\cite{rf:4} The result is given by

\begin{equation}
C_{m}^{0}={\rm const.}T^6,
\end{equation}
The temperature dependence of the emissivity can be understood easily. Let us investigate the quark URCA process (1). There are {\it u,d} quarks and electrons which are approximately degenerate with Fermi momentum $p_{F}(i)$. Each Fermion gives one power of $T$ from the phase space integral ($d^{3}p_i\to P_{F}(i)^{2}dE_{i}\propto T$). Thus we have $T^3$ from the {\it u,d} quarks and electrons. In addition, the phase space integral for the neutrino gives $d^{3}p_{\nu}\propto E_{\nu}^{2}dE_{\nu}\propto T^3$. One power of $T$ from the emitted neutrino energy, $E_{\nu}$, cancels a factor $T^{-1}$ from the energy-conserving $\delta$ function. Since the momenta of the degenerate particles are restricted to lie close to their respective Fermi surfaces, the angular integrals give no temperature dependence. Altogether, we thus have $C_{m}\propto T^6$. The constant on the right-hand side of the above equation depends on the more detailed calculation. Since the value is not important for our later discussions, we do not consider it.

This derivation of Eq.(17) is based on the assumption that each Fermi energies are larger than the temperature, $E_{F}\gg T$. Noting that $p_{F}\propto \rho^{1/3}$, this condition will be approximately satisfied for the high density quark matter near the critical temperature considered in this paper.

Our interest is now modification of the emissivity due to the color-superconductivity and its influence on the cooling behavior. Again see the weak decay (1). There are two kinds of {\it d} quarks: the (Cooper) paired particles and the unpaired ones. For the former case, the {\it d} quark might decay in the following mode:

\begin{equation}
(ud)\to u+u+e^{-}+\bar{\nu_e}.
\end{equation}
Noting that $\Delta_0\sim 100{\rm MeV}<T_c$, this process is prohibited by the energy conservation law. Therefore the $\beta$-decay is suppressed by the formation of the Cooper pairs. The rate is estimated to be 

\begin{equation}
r\simeq \frac{\rm the\,\, number\,\,of\,\, unpaired\,\, {\it d-}quark}{\rm total\,\, number\,\, of\,\,{\it d-}{\rm quark}}.
\end{equation}
Similarly the second process of (1),

\begin{equation}
(ud)+e^{-}\to d+d+\nu_{e},
\end{equation}
is suppressed in the same manner as the first one. Consequently the neutrino emissivity in super phase is given by

\begin{equation}
C_{em}=r^2 C_{em}^{(0)}.
\end{equation}
According to the usual BCS theory \cite{rf:9}, the rate of the unpaired {\it d} quark is given by the following equation:

\begin{equation}
r=\frac{1}{3}-\frac{2}{Tk_{F}^{4}}\int_{0}^{\infty}k^4 \frac{\partial
n_k}{\partial E_k}dk.
\end{equation}
The first term comes from the contribution of the gapless {\it d} quarks and the second from those of the unpaired {\it u,d} quarks. As a result, the neutrino emissivity is suppressed by the pairing correlation.

\section{Numerical results}
Numerical calculation has been carried out in the case of the chemical potential $\mu=500{\rm MeV}$. If the coupling constant $g$ is taken as $1{\rm GeV}$, the resulting gap energy becomes about $100{\rm MeV}$, which is a typical case used in other references. It is assumed that the emissivity (21) derived in the previous section is valid under these circumstances. 

First the behavior of the specific heat is drawn in Fig.1.
\begin{figure}[htb]
\parbox{\halftext}{
\epsfxsize=6.6cm
\epsfbox{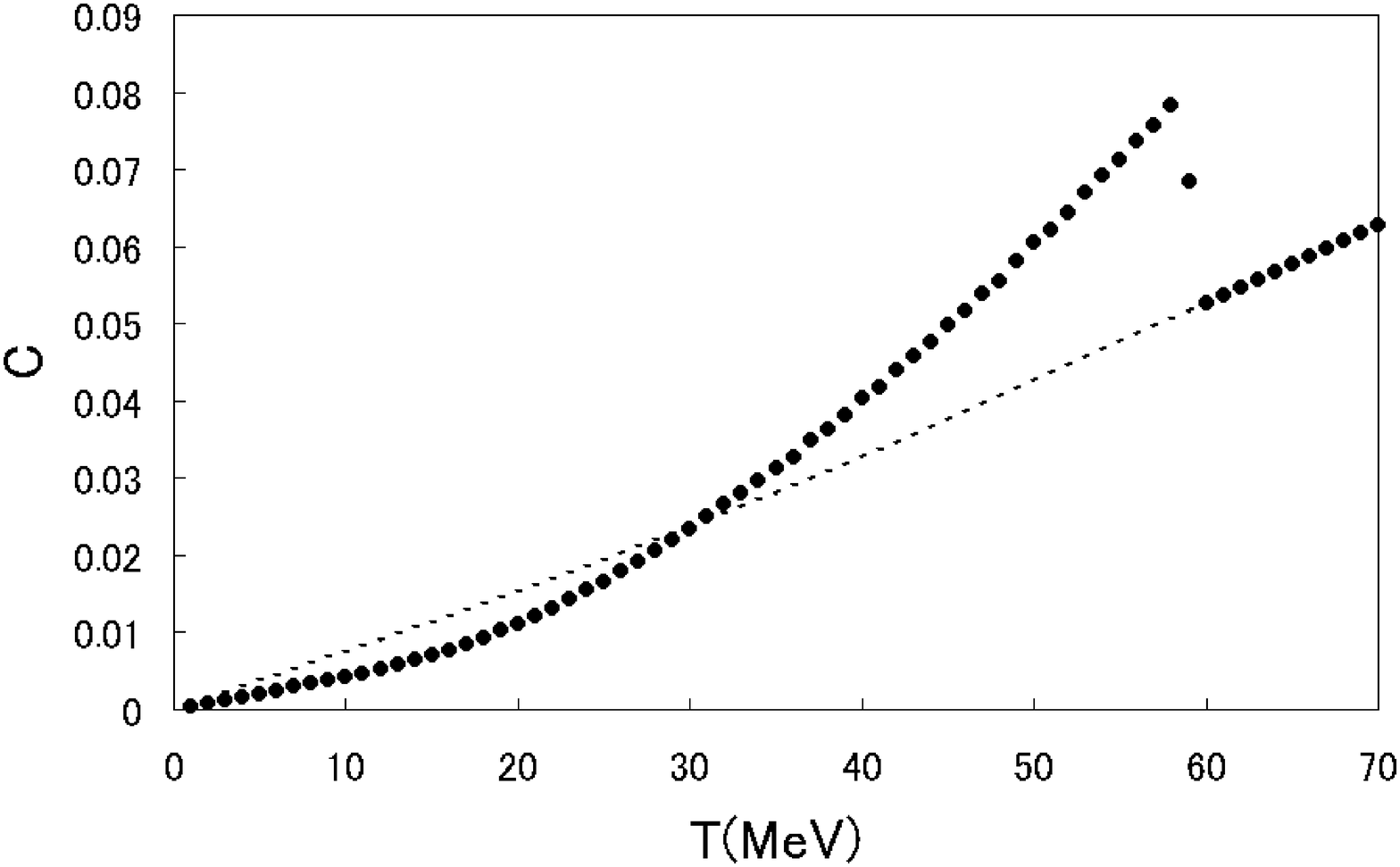}
\caption{The specific heat of quark matter as a function of temperature. The solid (dotted) line represents the result in the super (normal) phase.}}
\hspace{8mm}
\parbox{\halftext}{
\epsfxsize=6.6cm
\epsfbox{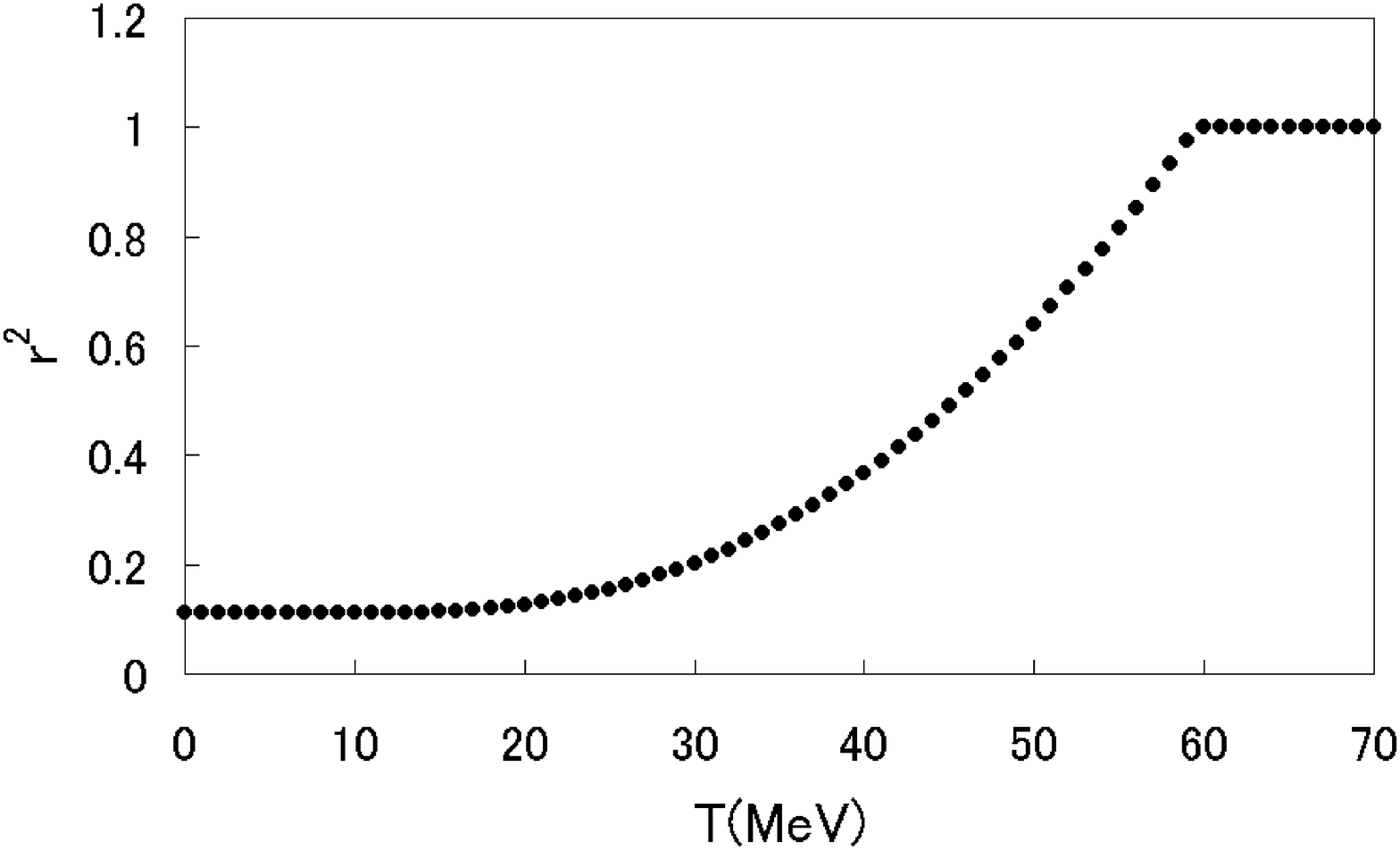}
\caption{The rate of unpaired {\it d}-quarks as a function of temperature. The rate is defined by the equation (22).}}
\end{figure}
The solid line represents the result in the super phase and dotted line denotes that in the normal one. We can see a discontinuity at the critical temperature representing the phase transition of second order. Below the critical temperature, the specific heat decreases with an exponential curve. On the other hand, it is linear in the normal phase so that the cooling rate in super phase becomes larger than that in normal phase at lower temperature. 

The rate of unpaired particles defined by Eq.(22) is shown in Fig.2. Since the Cooper pairs are formed in the super phase, the number of the unpaired quark diminishes below the critical temperature. Substituting the specific heat and emissivity considered above into Eq.(2), the cooling rate is calculated and the result is drawn in Fig.3 as a function of $T$. 
\begin{figure}[htb]
\epsfysize= 6.6cm
\centerline{\epsfbox{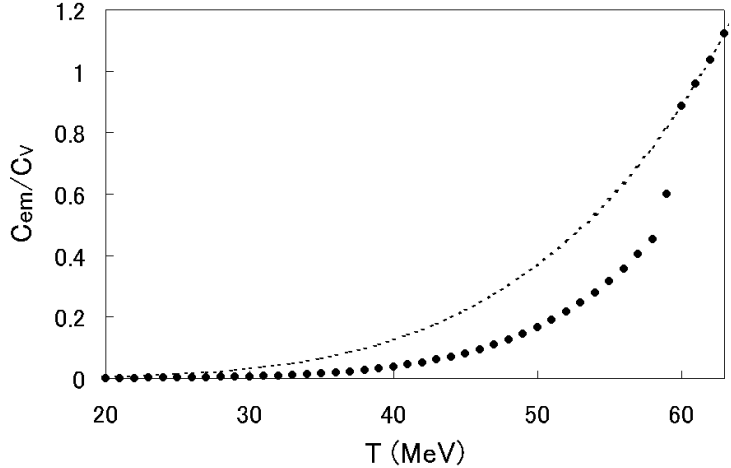}}
\caption{The cooling rate $-dT/dt=C_m(T)/C_V(T)$ as a function of temperature.}
\label{fig:3}
\end{figure}
We can see the pairing effects on the cooling rate as follows. First the sudden suppression of the cooling rate occurs across the critical temperature. This suppression is strengthened by the emissivity. Below the critical temperature, the decrease of the emissivity causes overall suppression of the cooling rate, in particular at lower temperature. 

Finally some comments are added as follows. The sudden suppression of the cooling rate may give a manifest signal for the color-superconductivity in quark matter if we could determine the temperature in the core of the stars in any way. However it is very difficult task at present.  Another possible signal is direct measurement of the decrease of the neutrino emissivity, such as measurement of the neutrino energy from a supernova. But this would be also very difficult. Anyway the observation of the signal is a future problem.

Second comment is about the other cooling process that gluons excited in hot quark matter decay into neutrino-antineutrino pairs:\cite{rf:8}

\begin{equation}
gluon\to q+\bar{q}\to Z_{0}\to \nu+\bar{\nu}.
\end{equation}
where $q+\bar{q}$ means a particle-hole state in the quark matter. This process seems to give large contribution to the cooling. In the super phase, however, the particle-hole state should be replaced by the two-quasi-particle state whose excitation energy is larger than $2\Delta_0$. Noting that the gluon energy is smaller than $T_c<2\Delta_0$, this process will be prohibited by the energy conservation law. Therefore this cooling process is suppressed by the formation of Cooper pairs and may be neglected in the discussions on the cooling behavior.

In conclusion, the neutrino emissivity decreases on account of the color superconductivity and causes suppression of the cooling rate. This phase transition leads to a sudden discontinuous suppression of the cooling rate in cooperation with the specific heat.

\section*{Acknowledgements}
We would like to thank Professor T.~Kunihiro at Kyoto University for the valuable comments and discussions.


\begin{thebibliography}{99}
\bibitem{rf:1}D.~Bailin and A.~Love, Phys.\ Rept.\ {\bf 107} (1984), 325, and references therein;\\
M.~Iwasaki and T.~Iwado, \PL{B350,1995,163};\\
M.~Alford, K.~Rajagopal and F.~Wilczek, \PL{B422,1998,247};\\
M.~Alford, K.~Rajagopal and F.~Wilczek, NP{B537,1998,443};\\
R.~Rapp, T.~Schaefer, E.V.~Shuryak and M.~Velkovsky, \PRL{81,1998,53}.
\bibitem{rf:2}K.~Rajagopal, hep-ph/0009058.
\bibitem{rf:3}M.~Iwasaki, T.~Tanaka and S.~Ishikawa, \PTP{104,2000,777}.
\bibitem{rf:4}N.~Iwamoto, \ANN{141,1982,1}, and references therein.
\bibitem{rf:5}M.~Alford, J.~Bowers and K.Rajagopal, J. Phys. G:Nucl. Par. Phys. {\bf 27} (2001), 541;\\
M.~Prakash, J.~Lattimer, J.~Pons, A.~Steiner and S.~Reddy, astro-ph/0012136;\\
D.~Page, M.~Prakash, J.~Lattimer and A.~Steiner, \PRL{85,2000,2048}.
\bibitem{rf:6}K.~Yamaguchi, Master thesis, Hiroshima University(1999);\\
R.~Horie, Master thesis, Kyoto University(1999).
\bibitem{rf:7}M.~Iwasaki, \PTP{101,1999,1043}.
\bibitem{rf:8}T.~Hatsuda, Mod. Phys. Lett. {\bf A2} (1987), 805;\\
T.~Hatsuda, C.S.~Lim,M and Yoshimura, Mod. Phys. Lett. {\bf A3} (1988), 1133;\\
C.~Schaab, B.~Hermann, F.~Weber and M.~Weigel, Astrophys. J. {\bf 480} (1997), L111;\\
J.D.~Anand, A,~Goyal, V.K.~Gupta and S.~Singh, Astrophys. J. {\bf 481} (1997), 954;\\
D.Blaschke, T.Kl{\"a}hn and D.N.Voskresensky, Astrophys. J. {\bf 533} (2000), 406.
\bibitem{rf:9}J.R.~Schrieffer, {\it Theory of Superconductivity} (Addison-Wesley, 1983) p.218.
\end{thebibliography}
\end{document}